\documentstyle [12pt] {article}
\title{THE SIMPLEST DETERMINATION OF THE THERMODYNAMICAL CHARACTERISTICS
OF SCHWARZSCHILD, KERR ANDREISSNER-NORDSTR$\ddot {\rm O}$M BLACK
HOLE}

\author{Vladan Pankovi\'c$^{\ast,\sharp}$,Jovan Ivanovi\'c$^\sharp$,\\
Milan Predojevi\'c $^\ast,^\sharp$, Aleksandar - Meda Radakovi\'c$^\sharp$  \\
$^\ast$Department of Physics, Faculty of Sciences, 21000 Novi
Sad,\\ Trg Dositeja Obradovi\'ca 4. , Serbia, vdpan@neobee.net \\
$^\sharp$Gimnazija, 22320 Indjija, Trg Slobode 2a, Serbia\\ }

\date {}
\begin{document}
\maketitle

\vspace {0.5cm}
\vspace {0.5cm}

\begin {abstract}
In this work, generalizing our previous results, we determine in
an original and the simplest way three most important
thermodynamical characteristics (Bekenstein-Hawking entropy,
Bekenstein quantization of the entropy or (outer) horizon surface
area and Hawking temperature) of  Schwarzschild, Kerr and
Reissner-Nordstr$\ddot {\rm o}$m black hole. We start physically
by assumption that circumference of black hole (outer) horizon
holds the natural (integer) number of corresponding reduced
Compton's wave length and use mathematically only simple algebraic
equations. (It is conceptually similar to Bohr's quantization
postulate in Bohr's atomic model interpreted by de Broglie
relation.)
\end {abstract}
\vspace {1.5cm}

In our previous work [1] we determined in an original and simplest
way important thermodynamical characteristics of  Schwarzschild
black hole. In this new work, generalizing our previous results,
we shall determine in analogous and the simplest way three most
important thermodynamical characteristics (Bekenstein-Hawking
entropy, Bekenstein quantization of the entropy or (outer) horizon
surface area and  Hawking temperature [2]-[8]) of  Schwarzschild,
Kerr and Reissner-Nordstr$\ddot {\rm o}$m black hole. We shall
start physically by assumption that circumference of black hole
(outer) horizon holds the natural (integer) number of
corresponding reduced Compton's wave length and use mathematically
only simple algebraic equations. (It is conceptually similar to
Bohr's quantization postulate in Bohr's atomic model interpreted
by de Broglie relation.)

Consider a black hole with mass $M$ and surface area of the outer
horizon
\begin {equation}
    A_{+}=8\pi R_{g}R_{+}= (8\pi (1+\sin\alpha) \frac {G^{2}M^{2}}{c^{4}})
\end {equation}
with radius
\begin {equation}
    R_{+}= R_{g} (1+\sin\alpha)
\end {equation}
where
\begin {equation}
     R_{g} = \frac {GM}{c^{2}}       .
\end {equation}

For
\begin {equation}
    \cos\alpha = \frac{1}{ R_{g}} \frac {J}{Mc}  = \frac {Jc}{M^{2}G}
\end {equation}
given black hole represents Kerr black hole with angular momentum
$J$. Especially for $J=0$ given black hole represents
Schwarzschild black hole.

For
\begin {equation}
    \cos\alpha = \frac{1}{R_{g}} \frac {Q}{c^{2}}(\frac {G}{4\pi \epsilon_{0}})^{\frac {1}{2}}  = \frac {Q}{M (G 4\pi \epsilon_{0})^{\frac
    {1}{2}}}
\end {equation}
given black hole represents Reissner-Nordstr$\ddot{\rm o}$m black
hole with electrical charge $Q$ where $\epsilon_{0}$ represents
vacuum dielectric constant. Especially, for $Q=0$ given black hole
represents Schwarzschild black hole.

Angular momentum $J$ represents a variable independent of $M$.
But, we can formally present $J$ as a function of $M$. Namely,
since $\cos\alpha$ must be smaller or equivalent to 1, $J$
according to (4) must satisfy the following inequality
\begin {equation}
   J \leq \frac {R_{g}}{Mc} = \frac {GM^{2}}{c}            .
\end {equation}
It admits that $J$ can be formally presented by
\begin {equation}
    J = R_{g}Mc \cos\alpha = GM^{2}\frac {\cos\alpha}{c}       .
\end {equation}
Given expression represents, of course, an identity $J=J$ but
formally (7) can be considered as a function of $M$ proportional
to $M^{2}$ where $\cos\alpha$ can be considered as a parameter.
Then formally differentiation of (7) yields
\begin {equation}
   dJ = 2GM \frac {\cos\alpha}{c}dM                 .
\end {equation}
As it will be shown later given formal procedure similarly to
renormalization procedure in quantum field theory, implies correct
physical results.

Analogously, electrical charge $Q$ represents a variable
independent of $M$. But, we can formally present $Q$ as a function
of $M$. Namely, since $\cos\alpha$ must be smaller or equivalent
to 1, $Q$, according to (5), must satisfy the following inequality
\begin {equation}
   Q \leq \frac {R_{g}}{(\frac {G}{ 4\pi \epsilon_{0}c^{4}})^{\frac {1}{2}}}= M(4\pi \epsilon_{0}G) ^{\frac {1}{2}}           .
\end {equation}
It admits that $Q^{2}$ can be formally presented by
\begin {equation}
    Q = R_{g}( \frac {4\pi \epsilon_{0}c^{4}}{G})^{\frac {1}{2}} \cos\alpha = M(4\pi \epsilon_{0} G)^{\frac {1}{2}}\cos\alpha
\end {equation}
that formally can be considered as a function of $M$ proportional
to $M$ where $\cos\alpha$ can be considered as a parameter. Then,
formally, differentiation of (10) yields
\begin {equation}
    dQ = (4\pi \epsilon_{0} G)^{\frac {1}{2}} \cos\alpha dM               .
\end {equation}
As it will be shown later, given formal procedure similarly to
renormalization procedure in quantum field theory, implies correct
physical results.

Suppose the following expression
\begin {equation}
      M_{+n}c R_{+} = n\frac {\hbar}{2\pi}, \hspace{0.5cm} {\rm for} \hspace{0.5cm}n = 1,
      2,...
\end {equation}
that implies
\begin {equation}
      2\pi R_{+} = n \frac {\hbar}{m_{+n}c}, \hspace{0.5cm} {\rm for} \hspace{0.5cm}n = 1, 2,...  .
\end {equation}
Here $2\pi R_{+}$ represents the circumference of the outer
horizon while
\begin {equation}
      \lambda_{r+n}= \frac {\hbar}{m_{+n}c}
\end {equation}
represents  $n-th$ reduced Compton wavelength of a quantum system
with mass $ m_{+n}$ captured at black hole outer horizon for $n =
1, 2,... …$ .

Expression (12) simply means that {\it circumference of the black
hole outer horizon holds exactly}$n$ {\it corresponding} $n$-{\it
th reduced Compton wave lengths of  quantum systems for} $n = 1,
2,... …$ . Obviously, it is conceptually similar to well-known
Bohr's quantization postulate interpreted by de Broglie relation
(according to which circumference of $n$-th electron circular
orbit contains exactly n corresponding $n$-th de Broglie wave
lengths, for $n = 1, 2,... …$).

According to (12) it follows
\begin {equation}
       m_{+n} = n\frac {\hbar}{2\pi c R_{+}}= n \frac {\hbar c}{2\pi GM(1+\sin\alpha)} \equiv n m_{+1}, \hspace{0.5cm} {\rm for} \hspace{0.5cm}n = 1,
       2,...
\end {equation}
where
\begin {equation}
       m_{+1} = \frac {\hbar c}{2\pi GM(1+\sin\alpha)}= \frac {M_{P}}{M}\frac{ M_{P}}{2\pi (1+\sin\alpha)}        .
\end {equation}
Obviously, $ m_{+1}$ depends of $M$ so that $ m_{+1}$ decreases
when $M$ increases and vice versa. For a macroscopic black hole,
i.e. for $M \gg M_{P}$ it follows $ m_{+1}\ll M_{P}$.

According to (16)
\begin {equation}
       \sigma_{+}= \frac {M}{ m_{+1}}= 2\pi (1+\sin\alpha) \frac {GM^{2}}{\hbar c}           .
\end {equation}
After multiplication of (17) by Boltzmann constant $k_{B}$, it
follows
\begin {equation}
       S_{BH+}= k_{B}\sigma_{+} = 2\pi (1+\sin\alpha) k_{B}\frac {GM^{2}}{\hbar c}= k_{B}\frac {A_{+}}{(2L_{P})^{2}}
\end {equation}
that represents Bekenstein-Hawking entropy $ S_{BH+}$ .

Differentiation of (18), under formal supposition that
$\sin\alpha$ does not represent a function of $M$ (even if,
exactly speaking, according to (4), (5), $\sin\alpha$ depends of
$M$) yields
\begin {equation}
    dS_{BH+}= k_{B}d\sigma_{+} = 4\pi (1+\sin\alpha) k_{B}\frac {GM}{\hbar c} dM
\end {equation}
or, in a corresponding finite difference form
\begin {equation}
   \Delta S_{BH+}= k_{B}\Delta \sigma_{+}= 4\pi (1+\sin\alpha) k_{B}\frac {GM}{\hbar c} \Delta M  \hspace{0.5cm} {\rm for} \hspace{0.5cm} \Delta M \ll M  .
\end {equation}
Further, we shall assume
\begin {equation}
   \Delta M = n m_{+1} \hspace{0.5cm} {\rm for} \hspace{0.5cm}n = 1,
   2,...
\end {equation}
which, after substituting in (20), yields
\begin {equation}
    \Delta S_{BH+} = k_{B}2n \hspace{0.5cm} {\rm for} \hspace{0.5cm}n = 1,
    2,...
\end {equation}
that represents Bekenstein quantization of the black hole entropy.
It, according to (18), implies
\begin {equation}
    \Delta A_{+} = 2n (2L_{P})^{2} \hspace{0.5cm} {\rm for} \hspace{0.5cm} n = 1,
    2,...
\end {equation}
that represents Bekenstein quantization of the black hole surface.

Now we shall consider first thermodynamical law for Kerr black
hole
\begin {equation}
   dE = T_{+}dS_{+} +  \Omega_{+}dJ
\end {equation}
where
\begin {equation}
   E=Mc^{2}
\end {equation}
represents the black hole energy, and,
\begin {equation}
    \Omega_{+} = \frac {c \cos\alpha}{2R_{g}(1+\sin\alpha)} = \frac {c^{3}\cos\alpha}{2MG(1+\sin\alpha)}
\end {equation}
- outer horizon rotation rate. Introduction of (25), (19), (26),
(8) in (24) yields
\begin {equation}
  c^{2}dM = T_{+}4\pi (1+\sin\alpha)k_{B}\frac {GM}{\hbar c}dM + \frac{c^{3}\cos\alpha}{2MG(1+\sin\alpha)}\frac {2GM \cos\alpha}{c}dM
\end {equation}
which, after simple calculations, yields
\begin {equation}
   T_{+} = \hbar c^{3}\frac {\sin\alpha}{4\pi (1+\sin\alpha) k_{B}GM}
\end {equation}
that represents Hawking temperature for Kerr black hole.

Now we shall consider first thermodynamical law for
Reissner-Nordstr$\ddot{o}$m black hole
\begin {equation}
    dE = T_{+}dS_{+} +  V_{+}dQ
\end {equation}
where
\begin {equation}
   V_{+}= \frac {Q}{4\pi \epsilon_{0}R_{+}}= \frac {Qc^{2}}{4\pi \epsilon_{0}MG(1+\sin\alpha)}
\end {equation}
represents the electrostatic potential at black hole outer
horizon. Introduction of (25), (19), (30), (11) in (29) yields
\begin {equation}
   c^{2}dM = T_{+}4\pi (1+\sin\alpha)k_{B}\frac{GM}{\hbar c}dM + \frac {Qc^{2}}{4 \pi \epsilon_{0}MG(1+\sin\alpha)}(( 4 \pi \epsilon_{0}G)^{\frac {1}{2}}\cos\alpha) dM
\end {equation}
which, after simple calculations, yields
\begin {equation}
    T_{+} = \hbar c^{3}\frac {\sin\alpha}{4\pi (1+\sin\alpha) k_{B}
GM} (1 -  \frac {Q \cos\alpha }{(4\pi \epsilon_{0}G)^{\frac
{1}{2}}M(1 + \sin\alpha)})
\end {equation}
that represents Hawking temperature for
Reissner-Nordstr$\ddot{o}$m black hole.

In this way we have reproduced, i.e. determined exactly, in the
simplest way, three most important thermodynamical characteristics
of Schwarzschild, Kerr and Reissner-Nordstr$\ddot{o}$m black hole:
Bekenstein-Hawking entropy (18), Bekenstein quantization of the
black hole entropy (22) or Bekenstein quantization of the black
hole surface area (23), and, Hawking temperature (28), (32).

\section {References}

\begin {itemize}

\item [[1]] V.Pankovic, M.Predojevic, P.Grujic, {\it A Bohr's Semiclassical Model of the Black Hole Thermodynamics}, gr-qc/0709.1812
\item [[2]] J. D. Bekenstein, Phys. Rev., {\bf D7}, (1973), 2333.
\item [[3]] S. W. Hawking, Comm. Math. Phys., {\bf 43}, (1975), 199
\item [[4]] S. W. Hawking, Phys. Rev., {\bf D14}, (1976), 2460.
\item [[5]] S. W. Hawking, in {\it General Relativity, an Einstein Centenary Survay}, eds. S. W. Hawking, W. Israel (Cambridge University Press, Cambridge, UK 1979)
\item [[6]] R. M. Wald, {\it Black Hole and Thermodynamics}, gr-qc/9702022
\item [[7]] R. M. Wald, {\it The Thermodynamics of Black Holes}, gr-qc/9912119
\item [[8]] D. N. Page, {\it Hawking Radiation and Black Hole Thermodynamics}, hep-th/0409024

\end {itemize}

\end {document}